# A new volatility term in the theory of options with transaction costs

Alexander Morozovsky[1]

Bridge, 57/58 Floors, 2 World Trade Center,

New York, NY 10048

E-mail: alex@nyc.bridge.com

Phone: (212) 390-6126

Fax: (212) 390-6498

The introduction of transaction costs into the theory of option pricing could lead not only to the change of return for options, but also to the change of the volatility. On the base of assumption of the portfolio analysis, a new equation for option pricing with transaction costs is derived. A new solution for the option price is obtained for the time close to expiration date.

---

[1] The ideas expressed in this article are the author's only and do not necessarily correspond to the views of Bridge.

## Introduction.

The problem of option pricing with transaction costs is one of the most interesting and important problems in the theory of option pricing [1,3]. The problem could be rephrased as a try to extend the results from Black-Scholes theory to the incomplete market. In the standard approach the possible return for the option is decremented by the transaction costs (the results derived by Leland and Wilmott [3] ). However, the introduction of transaction costs could also lead to the change of volatility. Because of this, new terms could appear in the option equation. On the base of this idea the new equation for option pricing with transaction costs is derived in this article. The new solutions are obtained and the conditions of their applicability are discussed.

## Method.

First of all, let us reproduce the derivation of Black - Scholes equation for call option on the base of risk-return argument (for simplicity's sake, we will consider the European call option). In order to do this we will write a change in security value in the specific form. We will assume that dP in time is determined by the sum of two factors: a factor that depends on the term proportional to dt and the factor proportional to dz:

$$dP = P_t * dt + P_z * dz, \qquad (1)$$

where t is time and z is Wiener process. Now we will use these notations in order to connect risk and return for different instruments. Proportionality of risk and return in these terms could be expressed in the form:

$$k = \frac{P_t - rP}{|P_z|} \qquad (2)$$

For option we could write $P_t$ as:

$$P_t = \frac{\partial P}{\partial t} + \mu S \frac{\partial P}{\partial S} + \frac{1}{2}\sigma^2 S^2 \frac{\partial^2 P}{\partial S^2} \qquad (3)$$

and $P_z$ :

$$P_z = \sigma S \frac{\partial P}{\partial S} \qquad (4)$$

Because the investor will make sure that relationship between risk and return for option and stock would be the same we could write that :

$$\lambda_{op} = \lambda_s ( \lambda_s = \frac{\mu - r}{\sigma} ) \qquad (5)$$

where $\lambda_{op}$ is the market price of risk for option and $\lambda_s$ is the market price of risk for stock. Combining (2) and (5) we could immediately obtain:

$$\frac{P_t - rP}{|P_z|} = \frac{\mu - r}{\sigma} \qquad (6)$$

Substituting $P_t$ and $|P_z|$ by (3) and (4) and considering that $\frac{\partial P}{\partial S} > 0$ for call options we immediately get:

$$\frac{\frac{\partial P}{\partial t} + \mu S \frac{\partial P}{\partial S} + \frac{1}{2}\sigma^2 S^2 \frac{\partial^2 P}{\partial S^2} - rP}{\sigma S \frac{\partial P}{\partial S}} = \frac{\mu - r}{\sigma} \qquad (7)$$

From (7) we receive usual Black - Scholes equation:

$$\frac{\partial P}{\partial t} + rS \frac{\partial P}{\partial S} + \frac{1}{2}\sigma^2 S^2 \frac{\partial^2 P}{\partial S^2} = rP \qquad (8)$$

Now, let's consider the same equation in the case of transaction costs. We will assume that we are using option for hedging and we are changing portfolio all the time. In this case: we need to find new changed factors from equation (2) - new return and new volatility. In order to do this let's rewrite differential dP into a new form:

$$dP = (\frac{\partial P}{\partial t} + \mu S \frac{\partial P}{\partial S} + \frac{1}{2}\sigma^2 S^2 \frac{\partial^2 P}{\partial S^2})dt - Ldt + (\sigma S \frac{\partial P}{\partial S}dz - A\,|dz| + Ldt), \qquad (9)$$

or

$$dP = dP_0 - Ldt + dP_1, \qquad (10)$$

where

$$dP_0 = (\frac{\partial P}{\partial t} + rS \frac{\partial P}{\partial S} + \frac{1}{2}\sigma^2 S^2 \frac{\partial^2 P}{\partial S^2})dt, \qquad (11)$$

$$dP_1 = \sigma S \frac{\partial P}{\partial S}dz - A\,|dz| + Ldt, \qquad (12)$$

$$A = \frac{1}{2}k\sigma S^2 \left|\frac{\partial^2 P}{\partial S^2}\right|, L = \frac{1}{2}k\sigma S^2 \left|\frac{\partial^2 P}{\partial S^2}\right|\sqrt{\frac{2}{\pi}}\frac{1}{\sqrt{dt}}, \quad (13)$$

and A, L – are terms connected with transactions cost.

From (13) one can immediately obtain that on average:

$$Ldt = A \ |dz|, \ <dP_1> = 0. \quad (14)$$

This means that the term $dP_1$ corresponds to volatility term in (2). When $dP_0$ - Ldt is proportional to the usual return for derivatives with transaction costs [3], $dP_1$ is constructed from different terms, one of which is the usual volatility term and the others appear because of transaction costs.

Let's calculate an average value of $dP_1^2$ <$dP_1^2$> :

$$<dP_1^2> = \sigma^2 S^2 \frac{\partial^2 P}{\partial S^2} dt + A^2 \ |dz|^2 + L^2 dt^2 - 2AL \ |dz| dt \quad (15)$$

because all terms, proportional to dz disappear (<dz> = 0). Let's remind also that:

$$<A|dz|> = Ldt \quad (16)$$

However, because $|dz|^2 = dz^2 \to dt$, we could rewrite (15) as:

$$<dP_1^2> = \sigma^2 S^2 \left(\frac{\partial P}{\partial S}\right)^2 dt + A^2 dt + \tilde{L}^2 dt - 2\tilde{L}^2 dt, \quad (17)$$

where $\tilde{L} = L\sqrt{dt}$.

This means that $<dP_1^2>$ is different from $\sigma^2 S^2 (\frac{\partial P}{\partial S})^2 dt$:

$$<dP_1^2> = \sigma^2 S^2 (\frac{\partial P}{\partial S})^2 dt + (A^2 - \tilde{L}^2) dt \qquad (18)$$

The origin of a new term in volatility is simple. Even, if $< A|dz| - Ldt > = 0$, it doesn't mean that $< (A|dz| - Ldt)^2 > = 0$.

From (18) and (14), (17) we immediately obtain:

$$<dP_1^2> = \sigma^2 S^2 (\frac{\partial P}{\partial S})^2 dt + (\frac{1}{2} k \sigma S^2 |\frac{\partial^2 P}{\partial S^2}|)^2 (1 - \frac{2}{\pi}) dt \qquad (19)$$

Because we could write $dP_0 - Ldt = P_t * dt$ and $dP_1^2 = P_z^2 * dt$, we could apply (6) for new $P_t$ and $P_z$:

$$\frac{P_t - L - rP}{P_z} = \frac{\mu - r}{\sigma} \qquad (20)$$

### Results

Now, we could rewrite (20) as:

$$\frac{\frac{\partial P}{\partial t} + \mu S \frac{\partial P}{\partial S} + \frac{1}{2} \sigma^2 S^2 \frac{\partial^2 P}{\partial S^2} - L - rP}{\sqrt{((\sigma S \frac{\partial P}{\partial S})^2 + (\frac{1}{2} k \sigma S^2 \frac{\partial^2 P}{\partial S^2})^2 (1 - \frac{2}{\pi}))}} = \frac{\mu - r}{\sigma} \qquad (21)$$

This a new equation for P (price of option with transaction costs). This equation confirms that the value of option price with transactions costs changed not only because transactions costs lead to changed return, but because transactions costs could also influence volatility. This equation equally shows the importance of both. Let us find the solution of this equation for two special cases:

1. Small transaction costs (k – small and we could consider $C = \frac{1}{2}k\sigma S^2 \left|\frac{\partial^2 P}{\partial S^2}\right|(1-\frac{2}{\pi})^{\frac{1}{2}}$ as much smaller term than $B = \sigma S \left|\frac{\partial P}{\partial S}\right|$:

$$C \ll B \qquad (22)$$

2. Price of option near time of expiration. In small interval close to time of expiration it is possible to assume [for example in [3], for the case of Asian options] that dominant term is proportional to $\frac{\partial^2 P}{\partial S^2}$ in comparison with $\frac{\partial P}{\partial S}$. Because of this we will consider situation when $B \ll C$, and denominator in (21) is mostly determined by the term proportional to $\left|\frac{\partial^2 P}{\partial S^2}\right|$.

**Small transactions costs.**

Substituting (22) into (21) we obtain:

$$\frac{\frac{\partial P}{\partial t} + \mu S \frac{\partial P}{\partial S} + \frac{1}{2}\sigma^2 S^2 \frac{\partial^2 P}{\partial S^2} - L - rP}{B(1+\frac{1}{2}\frac{C^2}{B^2})} = \frac{\mu - r}{\sigma} \qquad (23)$$

or

$$\frac{\partial P}{\partial t}+\mu S\frac{\partial P}{\partial S}+\frac{1}{2}\sigma^2 S^2\frac{\partial^2 P}{\partial S^2}-L-rP=(\mu-r)S/\frac{\partial P}{\partial S}/+\frac{1}{2}\frac{\mu-r}{\sigma^2 S/\frac{\partial P}{\partial S}/}(\frac{1}{2}k\sigma S^2\frac{\partial^2 P}{\partial S^2})^2(1-\frac{2}{\pi}) \quad (24)$$

The equation similar to well known equation ( [3] ) from the theory of options with transactions costs could be obtained from (24) in case we will completely neglect the last term D.

Than the equation (22) become:

$$\frac{\partial P}{\partial t}+\mu S\frac{\partial P}{\partial S}+\frac{1}{2}\sigma^2 S^2\frac{\partial^2 P}{\partial S^2}-L-rP=(\mu-r)S/\frac{\partial P}{\partial S}/ \quad (25)$$

This equation is similar to the new equation suggested in the paper [2], but with the influence of transactions costs (term L). Also it is usual equation from the theory of option pricing with transactions costs.

### Time close to the expiration.

If we completely neglect $\frac{\partial P}{\partial S}$ in the denominator of (21) we could get the following equation:

$$\frac{\partial P}{\partial t}+\mu S\frac{\partial P}{\partial S}+\frac{1}{2}\sigma^2 S^2\frac{\partial^2 P}{\partial S^2}-L-rP=\frac{1}{2}\frac{\mu-r}{\sigma}k\sigma S^2/\frac{\partial^2 P}{\partial S^2}/(1-\frac{2}{\pi})^{\frac{1}{2}} \quad (26)$$

This equation is very similar to the usual Black-Scholes equation, and if we assume that $\frac{\partial^2 P}{\partial S^2} > 0$ (what is correct for options without transactions costs), the equation (26) could be rewritten as:

$$\frac{\partial P}{\partial t} + \mu S \frac{\partial P}{\partial S} + \frac{1}{2} S^2 (\sigma^2 - \frac{\mu - r}{\sigma}(1 - \frac{2}{\pi})^{\frac{1}{2}} k\sigma) \frac{\partial^2 P}{\partial S^2} - L = rP \qquad (27)$$

The solution of this equation is Black-Scholes solution for the case of options with dividends, when we need to assume that one of parameters in this formula (rate of dividend) is negative:

$$q = -(\mu - r) \qquad (28)$$

and volatility parameter $\tilde{\sigma}$ is equal to:

$$\tilde{\sigma} = \sqrt{(\sigma^2 - (\mu - r)(1 - \frac{2}{\pi})^{\frac{1}{2}} k - k\sigma \sqrt{\frac{2}{\pi \delta t}})} \qquad (29)$$

The solution would be Black - Scholes solution:

$$P = SN(d_1) - Xe^{-(r-q)t} N(d_2), \qquad (30)$$

where volatility parameter used in this formula is $\tilde{\sigma}$.

# References.